\UseRawInputEncoding
\documentclass[showpacs,superscriptaddress,twocolumn]{revtex4-2}
\usepackage{bm,graphicx,dcolumn,epstopdf,epsf,latexsym,mathbbol,dcolumn,amssymb,amsmath,color,slashed,mathrsfs,mathcomp,simplewick}

\usepackage{multirow}
\usepackage{tabularx}
\usepackage{makecell}
\usepackage{color, soul}
\usepackage{hyperref}
\usepackage{url}

\usepackage{amssymb}
\usepackage{amsmath}
\usepackage{amsfonts}
\usepackage{slashed}
\usepackage{graphicx}
\usepackage{relsize}
\usepackage{color}
\usepackage{extarrows}
\usepackage{cancel}
\usepackage{placeins}
\usepackage{float}
\usepackage{subcaption}
\usepackage{caption}
\usepackage{here}
\usepackage[normalem]{ulem}

\begin{document}

\title{
Spinning Particles around Einstein-Geometric Proca AdS Compact Objects}

\author{Gulzoda Rakhimova}
\email{rakhimovagulzoda96@gmail.com}
\affiliation{Institute of Fundamental and Applied Research, National Research University TIIAME, Kori Niyoziy 39, Tashkent, 100000, Uzbekistan}
\affiliation{University of Tashkent for Applied Sciences, Str. Gavhar 1, Tashkent 100149, Uzbekistan}

\author{Beyhan Puli{\c c}e}
\email{beyhan.pulice@istinye.edu.tr}
\affiliation{Department of Basic Sciences, {\.I}stinye University, 34396, {\.I}stanbul, T\"{u}rkiye}
\affiliation{Faculty of Engineering and Natural Sciences, Sabanc{\i} University, 34956, {\.I}stanbul, T\"{u}rkiye}
\affiliation{Astrophysics Research Center,
The Open University of Israel, Raanana 4353701, Israel}

\author{Elham Ghorani}
\email{elham.ghorani@sabanciuniv.edu}
\affiliation{Faculty of Engineering and Natural Sciences, Sabanc{\i} University, 34956, {\.I}stanbul, T\"{u}rkiye}

\author{Farruh Atamurotov}
\email{atamurotov@yahoo.com}
\affiliation{Kimyo International University in Tashkent, Shota Rustaveli str. 156, Tashkent 100121, Uzbekistan}
\affiliation{Research Center of Astrophysics and Cosmology, Khazar University, 41 Mehseti Street, Baku AZ1096, Azerbaijan}

\author{Ahmadjon Abdujabbarov}
\email{ahmadjon@astrin.uz}
\affiliation{Institute of Fundamental and Applied Research, National Research University TIIAME, Kori Niyoziy 39, Tashkent, 100000, Uzbekistan}
\affiliation{School of Physics, Harbin Institute of Technology, Harbin 150001, People's Republic of China}
\affiliation{Andijan State University, Universitet Str. 129, Andijan 170100, Uzbekistan}
\affiliation{Tashkent State Technical University, Tashkent 100095, Uzbekistan}

\begin{abstract}
We investigate the dynamics of spinning test particles in the vicinity of Einstein--geometric Proca (EGP) Anti-de Sitter (AdS) compact objects, which arise from metric--Palatini gravity extended by the antisymmetric part of the affine curvature. Using the Mathisson--Papapetrou--Dixon (MPD) equations with the Tulczyjew spin supplementary condition, we derive the effective potential and analyze the equatorial motion of spinning particles. The influence of the model parameters $q_{1}$, $q_{2}$, and the Proca mass parameter $\sigma$ on the innermost stable circular orbits (ISCO), superluminal spin bounds, and orbital stability is systematically explored. Our results show that increasing $q_{1}$ and $q_{2}$ reduces the ISCO radius, angular momentum, and energy, while spin orientation introduces significant modifications to orbital behavior. We further examine head-on collisions of spinning particles near the horizon and demonstrate how the center-of-mass energy depends on spin and the EGP theory parameters. The study reveals that Einstein--geometric Proca AdS black holes may act as efficient particle accelerators, with distinctive features absent in Schwarzschild or standard AdS backgrounds. These findings provide new insights into the interplay between spin dynamics, modified gravity, and strong-field compact object physics.
\end{abstract}

\maketitle

\section{Introduction}

General relativity (GR) is usually derived from the Einstein-Hilbert action as a purely metric theory. However, it is well known that the variation principle for Einstein-Hilbert is not well-posed without adding the Gibbons-Hawking-York (GHY) boundary term \cite{york,gh}. This problem can be solved by replacing the Levi-Civita connection with an affine connection (independent of the metric) \cite{Palatini}. In this approach, known as the Palatini formalism, the Einstein-Hilbert action reproduces the Einstein field equations without the need for the boundary term. Palatini gravity is the simplest non-Riemannian extension of GR, which is characterized by metric $g_{\mu\nu}$ and the Ricci curvature ${\mathbb{R}}_{\mu\nu}(\Gamma)$ of a general symmetric affine connection $\Gamma^{\lambda}_{\mu\nu}$.

Palatini gravity naturally opens the door to non-Reimannian extensions of GR. A particularly important modification is obtained by adding a term like ${\mathbb{R}}_{[\mu\nu]}(\Gamma) {\mathbb{R}}^{[\mu\nu]}(\Gamma)$, where ${\mathbb{R}}_{[\mu\nu]}(\Gamma)$ is the anti-symmetric part of the affine Ricci tensor ${\mathbb{R}}_{\mu\nu}(\Gamma)$. The significance of this extension is that it leads to GR plus a purely geometric massive vector field $Q_{\mu}$ \cite{Vitagliano2010,Demir2020}. This field, which emerges directly from the affine connection, can be identified as a geometric Proca field. It is defined as
\begin{equation}
    Q_\mu \equiv \tfrac{1}{4} Q^\nu_{\ \mu\nu}, 
    \qquad 
    Q_{\lambda\mu\nu} \equiv - \nabla^\Gamma_\lambda g_{\mu\nu},
\end{equation}
where 
 $Q_{\lambda\mu\nu}$ is the non-metricity tensor \cite{Demir2020, Buchdahl1979,Tucker1996,Obukhov1997,Vitagliano2010}.

We refer to this system as Einstein-geometric Proca to distinguish it from the standard Einstein-Proca system studied in the literature. The latter describes general relativity minimally coupled to a matter Proca field, and has been analyzed extensively in various contexts such as the construction of Reissner-Nordstr\"{o}m (RN) type spherically symmetric vacuum solutions  \cite{Tresguerres1995a, 
Tucker1995, Vlachynsky1996, Macias1999}, the investigation of the physical 
role of the Proca field \cite{Bekenstein1971, Bekenstein1972, Adler1978}, the derivation of static spherically symmetric solutions \cite{Frolov1978, Gottlieb1984, Leaute1985}, and the study of horizon structure \cite{Ayon1999, Obukhov1999, Toussaint2000}. In contrast, the geometric Proca arises as a concequence of metric-incompatible symmetric connection and is in fact a distinctive feature of Weyl gravity\cite{Weyl0,Weyl1,Weyl2,Vitagliano2013}.  In cosmological settings, it has been studied as a 
candidate for geometric dark matter \cite{Demir2020}. Its couplings to fermions (quarks and leptons) were explored in \cite{dp-yeni} in regard to the black hole horizon in the presence of the Proca field \cite{Obukhov1997}.

Including the metrical curvature $R_{\mu\nu}({}^g\Gamma)$ in addition to the affine curvature 
${\mathbb{R}}_{\mu\nu}(\Gamma)$, relaxes the constraints on the Proca mass \cite{dp-yeni}. In the absence of the geometric Proca $Q_{\mu}$, this framework reduces to the usual metric-Palatini gravity, which has been explored in various domains, including dark matter \cite{Capozziello2012a}, wormholes \cite{Capozziello2012b}, and cosmology \cite{Capozziello2012c}. The theory we explore here is the metric-Palatini gravity extended with the invariant ${\mathbb{R}}_{[\mu\nu]}(\Gamma) {\mathbb{R}}^{[\mu\nu]}(\Gamma)$ and a negative cosmological constant (CC). This system is referred to as extended metric-Palatini gravity (EMPG). The black hole solution of this model was obtained and analyzed in \cite{AdS-Proca-1, AdS-Proca-2}. The thermodynamical analysis of the model was performed in \cite{Proca-Thermo}. It was shown in \cite{AdS-Proca-1} that the presence of a negative CC is essential for the existence of a black hole solution. The schematic action of EMPG is \cite{AdS-Proca-1}
 \begin{eqnarray}
\label{EMPG}
S[g,\Gamma]&=&\!\!\int\!\! d^4x \sqrt{-g} \Bigg \{\!``g^{\mu\nu}{R}_{\mu\nu}\left({}^g\Gamma\right)"+ ``g^{\mu\nu}{\mathbb{R}}_{\mu\nu}\left(\Gamma\right)"
 \nonumber \\ &&+ ``{\mathbb{R}}_{[\mu\nu]}(\Gamma) 
 {\mathbb{R}}^{[\mu\nu]}(\Gamma)" + ``{\rm CC}" \!\Bigg \},
\end{eqnarray} 
which describes an Einstein-geometric Proca Anti de Sitter (AdS) structure.

The study of particle motion in the vicinity of compact gravitating objects serves as an effective method to test different theories of gravity. Within the framework of metric theories, test particles are expected to follow geodesics, meaning that their trajectories encode information about the physical properties of the central source. However, particles endowed with intrinsic spin deviate from geodesics due to spin–curvature coupling. Investigating the dynamics of such spinning particles thus offers deeper insight into the geometry of spacetime. Various analyzes of neutral, charged, and magnetized particle dynamics around compact objects have been reported in the literature~\cite{Banados09,2018PhRvD..97j4024D,2013PhRvD..88b4016S,2011PhRvD..83b4016F,2021PhRvD.103h4057B,2011PhRvD..83d4053A,2004CQGra..21..961D,2023EPJP..138..846N,2020EPJC...80..399H,2022EPJP..137..634A,2022EPJC...82..659A}. Furthermore, the electromagnetic field surrounding compact objects can strongly influence the trajectories of charged or magnetized particles~\cite{2002MNRAS.336..241A,2015EPJC...75...24J,2020PhRvD.102d4013N,2014PhRvD..90d4029K}. The role of spin interactions with the background geometry has also been extensively addressed in earlier works~\cite{2019PhRvD.100j4052T,2020PhRvD.101b4037Z,2019PhRvD..99j4059C,2021PhRvD.104h4024B,2022PhRvD.106b4012A,2022IJMPD..3150091L,2014PhRvD..90f4035H,Abdulxamidov:2023jfq,Abdulxamidov:2023uvo,2015PhRvD..91l4030J,2021PhRvD.104b4042T,Jumaniyozov:2025nnj,Oteev:2025yjh,2023PhRvD.108d4041S,2023EPJP..138..245G}.

The behavior of spinning particles is commonly described by the Mathisson-Papapetrou-Dixon (MPD) equations~\cite{2011CQGra..28s5025P}, which incorporate the interaction between the spin tensor and the spacetime curvature. A characteristic feature of this framework is spin precession~\cite{Mathisson:1937zz,Papapetrou:1951pa,Corinaldesi:1951pb}. These dynamical processes can also influence the energetic aspects of black holes (BHs). In particular, BHs may act as natural particle accelerators. Banados, Silk, and West demonstrated in~\cite{BSW2009} that, under specific parameter conditions and in the case of extremal rotating BHs, the center-of-mass energy of two colliding particles can grow without bound—a phenomenon now known as the BSW mechanism.  In this work, we aim to investigate the collisions of spinning particles near the compact objects within the EMPG model. The impact of modified gravity scenarios on spin dynamics has been addressed in~\cite{2022PhRvD.105l4036G,2023EPJC...83.1031S,2023EPJP..138..635A,2023arXiv230212352L}. Moreover, several recent studies have explored spinning particle dynamics within diverse modified gravity frameworks. In particular, the effects of spin–curvature coupling and quantum corrections have been examined for Reissner–Nordström-like black holes \cite{Abdukayumova:2025ztr}, effective quantum gravity models \cite{Umarov:2025wzm}, quantum-improved charged black \cite{Ladino:2023zdn} and quantum-corrected spacetimes \cite{Alimova:2025izs}. Further analyses have addressed the influence of asymptotically safe gravity \cite{Mannobova:2025uqf}, Lorentz gauge theory \cite{Umarov:2025ihy}, T-duality–inspired charged geometries \cite{Rakhimova:2024ind}, and decoupled hairy black holes \cite{Rakhimova:2024hzt}. These works collectively enrich the theoretical landscape of spinning test-particle motion in alternative gravitational settings. 

In the present work, we study the spinning particles dynamics around the Einstein-geometric Proca AdS compact objects. The paper is organized as follows. We will describe the static spherically symmetric solution of EMPG in Section \ref{sect:2}. In Section \ref{sec3}, the equations of motion for a spinning particle is obtained and the superluminal bound is studied. We discuss the dynamics of spinning particle in Section \ref{sec4} and we finish the paper by concluding remarks in Section \ref{sec5}.

\section{Static Spherically-Symmetric Solutions in EMPG Model\label{sect:2}}

The exact form of the schematic action \ref{EMPG} is given by
\cite{Demir2020,dp-yeni,AdS-Proca-1}
 \begin{eqnarray}
S[g,\Gamma]&=&\int d^4x \sqrt{-g} \Bigg \{
\frac{M^2}{2} {R}\left(g\right) + \frac{{\overline{M}}^2}{2} {\mathbb{R}}\left(g,\Gamma\right) 
\nonumber \\&&+ \xi {\overline{\mathbb{R}}}_{\mu\nu}\left(\Gamma\right) {\overline{\mathbb{R}}}^{\mu\nu}\left(\Gamma\right) -V_0 + 
{\mathcal{L}}_{m}({}^g\Gamma,\psi) \Bigg \}, 
\label{mag-action}
\end{eqnarray}
which is an action based on the metric $g_{\mu\nu}$ and a symmetric affine connection $\Gamma^\lambda_{\mu\nu}=\Gamma^\lambda_{\nu\mu}$, independent of the Levi-Civita connection
\begin{eqnarray}
{}^g\Gamma^\lambda_{\mu\nu} = \frac{1}{2} g^{\lambda \rho} \left( \partial_\mu g_{\nu\rho} + \partial_{\nu} g_{\rho\mu}-\partial_\rho g_{\mu\nu}\right)
\end{eqnarray}
generated by the metric $g_{\mu\nu}$. The affine connection defines the affine Riemann curvature
\begin{eqnarray}
\label{affine-Riemann}
{\mathbb{R}}^\mu_{\alpha\nu\beta}\left(\Gamma\right) = \partial_\nu \Gamma^\mu_{\beta\alpha} - \partial_\beta \Gamma^\mu_{\nu\alpha} + \Gamma^\mu_{\nu\lambda} \Gamma^\lambda_{\beta\alpha} -\Gamma^\mu_{\beta\lambda} \Gamma^\lambda_{\nu\alpha}.
\end{eqnarray} 
The antisymmetric Ricci tensor in \ref{mag-action} is obtained by the contraction ${\overline{\mathbb{R}}_{\mu\nu}}\left(\Gamma\right) = {\mathbb{R}}^\lambda_{\lambda\mu\nu}\left(\Gamma\right) = {\mathbb{R}}_{[\mu\nu]}\left(\Gamma\right)$, which vanishes in the metrical geometry.

The first term in the action (\ref{mag-action}) presents the Einstein-Hilbert term in GR, provided that $M$ is identified with the Planck mass $M_P$. The second term corresponds to the standard Platini action. The third term, proportional to $\xi$, extends the metric--Palatini gravity 
with the antisymmetric part of the affine Ricci tensor \cite{Demir2020,dp-yeni}. Here, $V_0$ denotes the vacuum energy density, and ${\mathcal{L}}_{m}({}^g\Gamma,\psi) $ is the Lagrangian of the matter fields $\psi$. 

Since the affine connection $\Gamma^\lambda_{\mu\nu} = \Gamma^\lambda_{\nu\mu}$ 
is torsion-free, the only possible deviation from the Levi--Civita connection 
${}^{g}\Gamma^\lambda_{\mu\nu}$ is due to non-metricity, 
\begin{align}\Gamma^\lambda_{\mu\nu}= {}^g\Gamma^\lambda_{\mu\nu} +  \frac{1}{2} g^{\lambda \rho} ( Q_{\mu \nu \rho } + Q_{\nu \mu \rho } - Q_{\rho \mu \nu} ),
\label{fark-connection}
\end{align}
where the non-metricity tensor is defined as
\begin{equation}
    Q_{\lambda \mu \nu} = - {}^\Gamma \nabla_{\lambda} g_{\mu \nu}.
\end{equation}
Using the decomposition \ref{fark-connection} in the metric-Palatini action (\ref{mag-action}) leads to the reduced  Einstein-geometric Proca action \cite{Demir2020, dp-yeni,AdS-Proca-1}
\begin{eqnarray}\label{action-reduced}
S[g,Y,\psi] &=& \int d^4 x \sqrt{-g} \Bigg\{\frac{1}{16 \pi G_N} R(g)-V_0 - \frac{1}{4} Y_{\mu \nu} Y^{\mu \nu} \nonumber \\&& - \frac{1}{2} M_Y^2 Y_{\mu} Y^{\mu}+{\mathcal{L}}_{m} (g,{}^g \Gamma,\psi) \Bigg \}.
\end{eqnarray}
Here, $Q_\mu = Q _{\mu \nu}^{\nu}/4$ is the non-metricity vector,  $Y_\mu=2 \sqrt{\xi} Q_\mu$ is the canonical geometric Proca field, $G_N=8\pi/(M^2 + \overline{M}^2)$ is Newton's gravitational constant, and
\begin{eqnarray}
M_Y^2 = \frac{3 \overline{M}^2 }{2 \xi}
\end{eqnarray}
is the squared mass of the $Y_\mu$. It is convenient to define $\hat{Y}_\mu \equiv \sqrt{\kappa} Y_{\mu}$ as the canonical dimensionless Proca field and rewrite the reduced action (\ref{action-reduced}) 
 \begin{eqnarray}\label{action-EGP}
     S[g,Y] &=& \int d^4 x \sqrt{-g} \frac{1}{2 \kappa} \nonumber \\&& \times \Bigg\{R(g) - 2 \Lambda  -  M_Y^2 \hat{Y}_{\mu} \hat{Y}^{\mu}
- \frac{1}{2} \hat{Y}_{\mu \nu} \hat{Y}^{\mu \nu} \Bigg\}
 \end{eqnarray}
in which $\kappa = 8 \pi G_N$ and $\Lambda=8\pi G_N V_0$ is the CC. The variation of \ref{action-EGP} with respect to  $g_{\mu\nu}$ leads to the Einstein field equation
\begin{eqnarray}\label{Einstein-eqns}\nonumber
   R_{\mu \nu} - \Lambda g_{\mu \nu} - \hat{Y}_{\alpha\mu} \hat{Y}^{\alpha}_{\;\;\;\nu}  + \frac{1}{4} \hat{Y}_{\alpha \beta} \hat{Y}^{\alpha \beta} g_{\mu \nu} - M_Y^2 \hat{Y}_{\mu} \hat{Y}_{\nu} = 0, 
\end{eqnarray}
and variation of the action with respect to $\hat{Y}_\mu$ gives us the Proca equation of motion
\begin{align}
\nabla_\mu \hat{Y}^{\mu \nu} - M^2_Y \hat{Y}^\nu = 0.
\label{eom-Y}
\end{align}
This system of equations has been thoroughly studied in the search for black
hole solutions with $\Lambda = 0$ \cite{dp-yeni} and $\Lambda < 0$ \cite{AdS-Proca-1}. Putting forward a general spherically-symmetric and static metric ansatz
\begin{align}
g_{\mu \nu} = \text{diag}(-h(r),\frac{1}{f(r)} ,r^2, r^2 \sin ^2 \theta),
\label{metric-sss}
\end{align}
together with a time-like Proca field 
\begin{align}
\hat{Y}_\mu = \hat{\phi}(r) \delta_\mu^0, 
\label{proca-sss}
\end{align} 
one arrives at the following solution
\begin{equation}
\hat{\phi}(\hat{r})=\frac{q_1}{\hat{r}^{\frac{1-\sigma}{2}}} + \frac{q_2}{\hat{r}^{\frac{1+\sigma}{2}}}, 
\label{psi}
\end{equation}
 where $\sigma$ is the mass parameter
\begin{align}\label{sigma}
\sigma=\sqrt{1 + 4 \hat{M}_Y^2 l^2} ~    
\end{align}
and $l$ stands for the AdS radius.
According to the Breitenlohner--Freedman mass bound, $\sigma$ must satisfy the range \cite{Einstein-Proca}
\begin{align}
0 \leq \sigma < 1,
\label{sigma-range}
\end{align}
in order to prevent tachyonic runaway instabilities in the AdS background. 
Here, $q_1$ and $q_2$ are integration constants, interpreted as a uniform potential and an electromagnetic-like charge, respectively. $\hat{r}$ and ${\hat M}_Y$ are dimensionless distance and mass defined as
\begin{align}
\hat{r} := \kappa^{-1/2} r,~ 
{\hat M}_Y^2 := \kappa M_Y^2~
\end{align}
 Corresponding to the geometric Proca solution in (\ref{psi}), the metric potentials $f(r)$ and $h(r)$ are modified as \cite{AdS-Proca-1}
\begin{align}
\label{metric-funcs-param}
f(\hat{r}) &= \hat{r}^2l^{-2} + 1 + \frac{n_1}{\hat{r}^{1 - \sigma}} + \frac{n_2}{\hat{r}}\ ,\nonumber\\
h(\hat{r}) &= \hat{r}^2l^{-2} + 1 +\frac{m_1}{\hat{r}^{1 - \sigma}}+\frac{m_2}{\hat{r}}
\end{align}
where the explicit forms of the constants depend on the model parameters 
$(q_1,q_2,\sigma,l)$,
\begin{align}
\label{solution}
&n_1 = \frac{1 - \sigma}{4} q_1^2\ , \quad m_1 = \frac{1 - \sigma}{3 - \sigma} q_1^2\ , \nonumber \\
&n_2 = m_2 - \frac{(1 - \sigma)(1 + \sigma)}{6} q_1 q_2 \ . 
\end{align}
By setting $q_1=0$, we will recover the Ads-Schwarzschild solution.
The ADM mass of this compact object takes the form \cite{AdS-Proca-1}
\begin{eqnarray}
\label{mass}
    M_{ADM}=\frac{1}{2}\left(q_1 q_2 \left[\gamma  \sigma +\frac{1}{3} (1-\sigma ) (\sigma +4)\right]-m_2\right) \, .
\end{eqnarray}
where $\gamma$ is the coefficient of the surface term for the geometric Proca 
after the normalization $M_{ADM}=1$. This Einstein-geometric Proca AdS solution has been extensively 
analyzed, including the horizon structure, the structure of the singularity 
\cite{AdS-Proca-1} and the thermodynamics features \cite{AdS-Proca-Thermo}. 

\begin{figure}[t]

    \includegraphics[scale=0.65]{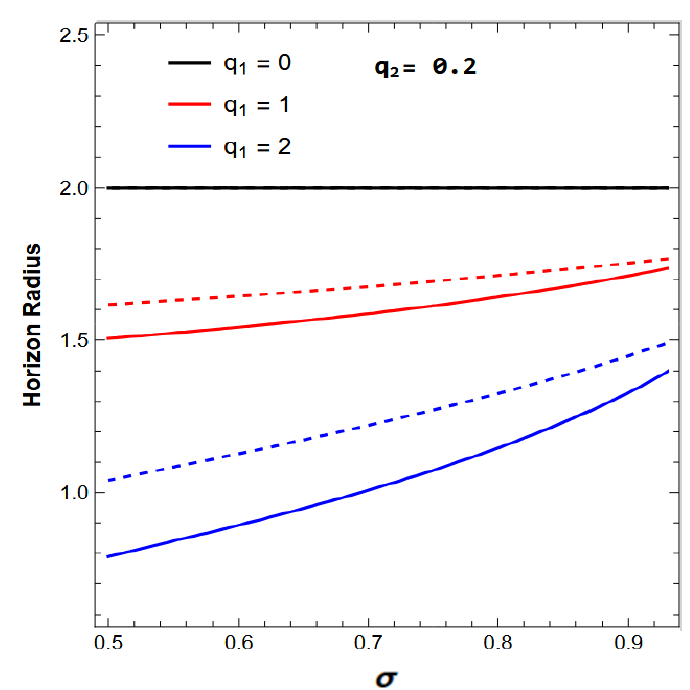}
    
 \caption{Killing horizon (solid) and event horizon (dashed) for various values  of the charge parameter $q_1$ and for fixed $q_2$ as a function of $\sigma$.} 
    \label{horizons}

\end{figure}

 The event horizon is located at values of $r$ for which $f(r)=0$ while $h(r)=0$ gives us the killing horizon. We solve the horizon radii numerically from the metric functions (\ref{metric-funcs-param}), and illustrate both horizons as functions of $\sigma$ in Fig. (\ref{horizons}). The solid lines represent the Killing horizon for various values of the charge parameter $q_1$ and for fixed $q_2$, while the dashed lines show the corresponding event horizons for the same configurations.
At fixed $\sigma$, increasing $q_1$ reduces the horizon radii, showing that a stronger Proca charge yields a more compact geometry.
As expected, the Killing horizon is always hidden behind the event horizon. The difference between these two horizons increases with the Proca charge $q_1$, while in the case $q_1 = 0$ the Killing and event horizons coincide at $r=2$; consistent with standard Schwarzschild solutions. The figure shows that both horizons decrease with higher Proca charge $q_1$, while they increase with $\sigma$. 

\section{Equations of motion: Basic concepts
\label{sec3}}
\subsection{Equations of motion for a spinning particle \label{sec3a}}
In this section, we consider some theoretical background of a spinning particle motion around compact objects. It is well-known that a non-spinning particle follows usual geodesic equations, whereas  in the case of the spinning particles the scenario is not simple. The set of equations used to describe the motion of a massive spinning particle in a gravitational field is called the Mathisson-Papapetrou-Dixon (MPD) equations, which are given in the form ~\cite{Mathisson:1937zz,Papapetrou:1951pa}: 
\begin{equation}
 \label{s3e1}
\begin{aligned}
\frac{Dp^\alpha}{d\lambda}&=-\frac{1}{2}R^\alpha_{\;\;\beta\delta\sigma}u^\beta S^{\delta\sigma},\\
\frac{D S^{\alpha\beta}}{d\lambda}&=p^\alpha u^\beta-p^\beta     u^\alpha
    \end{aligned}
    \end{equation}
where $D/d\lambda\equiv u^\alpha\nabla_\alpha$ is the projection of the covariant derivative along the trajectory of the particle, $u^\mu=dx^\mu/d\lambda$ is the 4-velocity of the test particle, $p^\alpha$ is the canonical 4-momentum, $R^\alpha_{\;\;\beta\delta\sigma}$ is the Riemann curvature tensor,  $\lambda$ is an affine parameter and $S^{\alpha\beta}$  is antisymmetric spin tensor:  $S^{\alpha\beta}=-S^{\beta\alpha}$.

Obviously, Eq.~(\ref{s3e1}) returns to the well-known geodesic equation of GR when the components of $S^{\alpha\beta}$ vanish, and consequently, the differential equation will take the form of
 \begin{equation}
    \label{s3e3}
    \frac{Dp^\alpha}{d\lambda}=0\ . 
    \end{equation} 
To solve the set of Eq.~(\ref{s3e1}), one needs to introduce the extra condition, the so-called spin-supplementary condition, which fixes the center of mass of the particle. There are several different spin-supplementary conditions to choose (e.g.~\cite{tulczyjew1959motion,1998PhRvD..58f4005S,Pirani:1956tn}). In this paper, we employ the Tulczyjew Spin Supplementary Condition (SSC)~\cite{tulczyjew1959motion}, given by the relation:
    \begin{equation}
    \label{s3e6}
    S^{\alpha\beta}p_\alpha=0.
    \end{equation}
The SSC relation and the MPD equations provide two independent conserved quantities: the canonical momentum and the particle's spin, given by the relations
\begin{equation}
    \label{s3e7}
\begin{aligned}
    S^{\alpha\beta}S_{\alpha\beta}&=2S^2=2m^2s^2,\\
    p^\alpha p_\alpha&=-m^2.
\end{aligned}
\end{equation}
In addition to the spin and the canonical momentum shown in Eq.~(\ref{s3e7}), one also has the usual conserved quantities associated with the spacetime symmetries. In the case of an axially symmetric spacetime, there are two Killing vector fields. One is related to invariant time translations, $\xi^\alpha$, and the other generates rotations along the azimuth angle $\phi$, $\psi^\alpha$. We can compute these quantities using the following equation:
    \begin{equation}
    \label{s3e8}
    p^\alpha \kappa_\alpha-\frac{1}{2}S^{\alpha\beta}\nabla_\beta\kappa_\alpha=p^\alpha \kappa_\alpha-\frac{1}{2}S^{\alpha\beta}\partial_\beta\kappa_\alpha=\text{constant},
    \end{equation}
where $k^\alpha$ represents one of the two Killing vector fields: $\xi^\alpha$ or $\psi^\alpha$.
 
\subsection{Superluminal bound
\label{sec3b}}
Considering the superluminal bound part while investigating the motion of a spinning particle around compact objects plays a crucial role. The spinning particle cannot have any spin value $s$ as its velocity can exceed the speed of light. Due to this, the critical value for a particle's spin should be found to keep its trajectory time-like and work with the data that has  physical meaning.

Obviously, four-momentum $p^\alpha$ is a conserved quantity (see, e.g., Eq. (\ref{s3e7}))  and always time-like while the four-velocity of the particle $u^\alpha$ is not. Therefore, the normalization condition $u^\alpha u_\alpha=-1$ is not adequately satisfied, and for some values of spin $s$, the values of the normalization condition can be positive. Thus, the trajectory of the particle becomes space-like, without physical meaning.

\begin{figure*}[t]
    \begin{center}
    \includegraphics[scale=0.6]{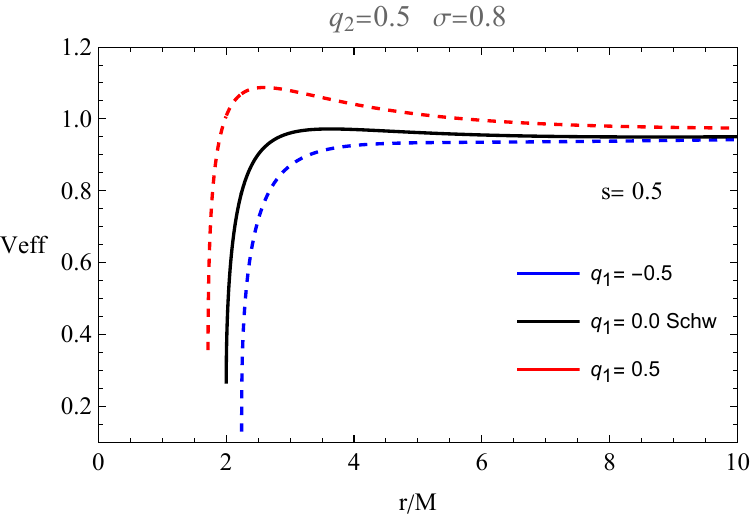}
    \includegraphics[scale=0.6]{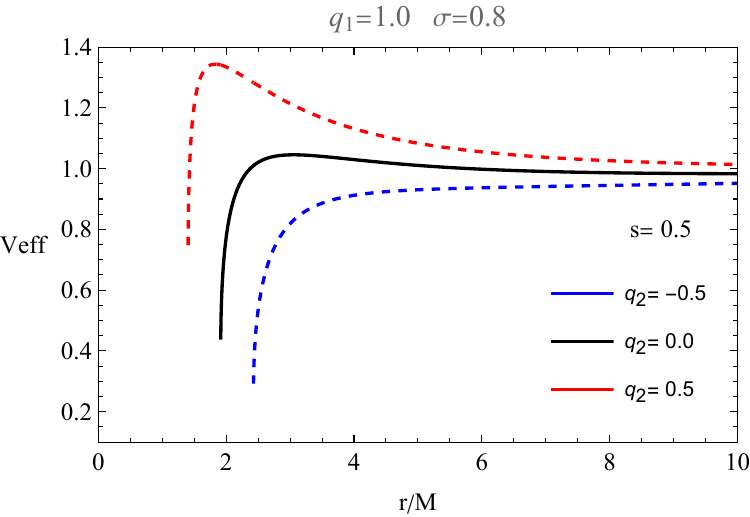}
    \includegraphics[scale=0.6]{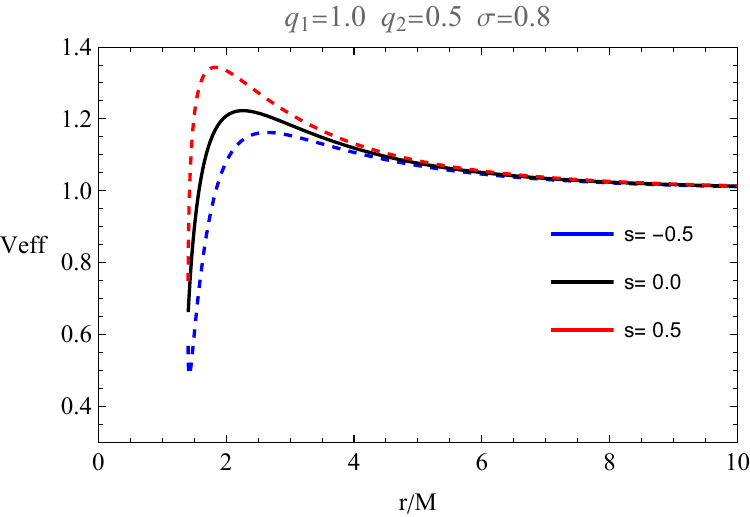}
    \includegraphics[scale=0.6]{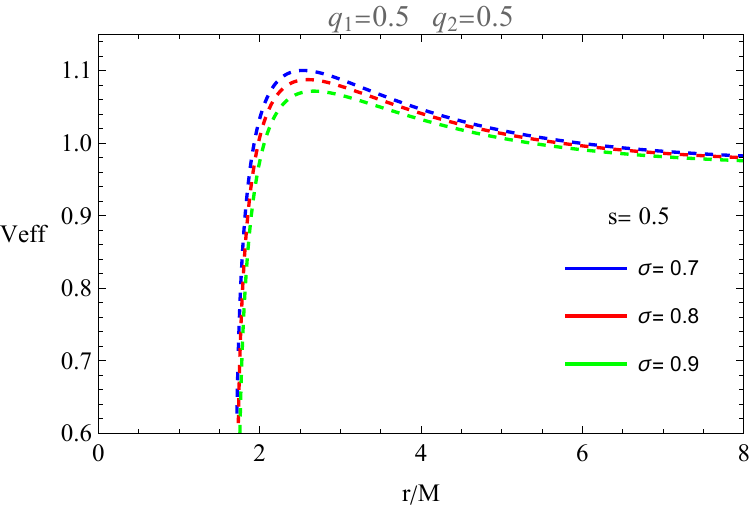}
    
 \caption{Radial dependence of the effective potential for different values of parameters. The upper line corresponds to the plots of the effective potential for various values of $q_1$ when  $q_2=0.5$ (left panel) and for various values of $q_2$ when $q_1=1.0$ (right panel) for  $s=0.5$ and $\sigma=0.8$. While the lower line illustrates the behavior of $V_{eff}$ for various values of $s$ for $q_1=1.0$ and $q_2=0.5$ (left panel) and for various values of $\sigma$ for $q_1,q_2,s=0.5$ (right panel)}.  
    \label{veffs}
    \end{center}
\end{figure*}

To maintain the trajectory of spinning test particles with time-like character, it is necessary to impose the following constraint (on the equatorial plane): %
    \begin{equation}
    \label{s3be1}
    \frac{u_\alpha u^\alpha}{(u^t)^2}=g_{tt}+g_{rr}\dot{r}^2+g_{\varphi\varphi}\dot{\varphi}^2\leq 0.
    \end{equation}
Here, the dot denotes the derivative with respect to the time coordinate $t$. Solving the MPD equations (see Eq. \ref{s3e1}), one gets the expressions for $dr/dt$ and $d\varphi/dt$ in the following form:
\\
   \begin{equation}
    \label{s3be10}
    \begin{aligned}
    \frac{dr}{dt}=\frac{u^r}{u^t}&=\frac{\mathcal{C}p_r}{\mathcal{B}p_t},\\\\
    \frac{d\varphi}{dt}=\frac{u^\varphi}{u^t}&=\frac{\mathcal{A}p_\varphi}{\mathcal{B}p_t}.
    \end{aligned}
    \end{equation}
    where
    \begin{equation}
    \label{s3be9}
    \begin{aligned}
    \mathcal{A}&=g^{\varphi\varphi}+\left(\frac{S^{\varphi r}}{p_t}\right)^2R_{trrt},\\\\
    \mathcal{B}&=g^{tt}+\left(\frac{S^{\varphi r}}{p_t}\right)^2R_{\varphi rr \varphi},\\\\
    \mathcal{C}&=g^{rr}+\left(\frac{S^{\varphi r}}{p_t}\right)^2R_{\varphi tt \varphi}.\\\\
    \end{aligned}
    \end{equation}

As have already been pointed out, Eq.~\ref{s3be1} is necessary to find the superluminal boundary, which helps to separate time-like particles from space-like ones. So, by defining the function $\mathcal{F}=u_\alpha u^\alpha/(u^t)^2$,  we can find the limiting values of the particle's spin. Obviously, the condition $\mathcal{F}=0$ corresponds to the superluminal constraint, the critical value of particle's spin $s$ for a particular condition. Consequently, when $\mathcal{F}<0$ for certain values of the spin $s$, the trajectory of the particle is time-like and the spin values satifying the condition are acceptable. However, when $\mathcal{F}>0$  the values of the particle's spin $s$ are forbidden and the trajectory of the particle becomes space-like. Note, that the behaviour of the limiting values of particle's spin $s_{max}$ is presented in the Fig.~\ref{scr} and described in the next section together with the innermost stable circular orbits due to their interconnectedness.

\section{Dynamics of spinning 
particles around Einstein-Geometric Proca AdS Compact Objects 
\label{sec4}}

 In this section we are aimed to investigate the motion of a spinning particle around Einstein Geometric-Proca  AdS compact object. In order to find the equations of motion for a spinning particle we use the effective potential method and investigate its circular orbits.   

\subsection{The Effective potential\label{sec4a}}
First, we intend to find the effective potential of the spinning test particle around Einstein-Geometric Proca compact object. For simplicity, we consider the motion on the equatorial plane ($\theta=\pi/2$).
%
There are two conserved quantities in the case of static spherically symmetric spacetimes, namely, the energy $E$ and the total angular momentum $J$ ($J=L+S$, $S=sm$, $L=\mathcal{L}m$) that are defined by the relations:
\begin{equation}
        \label{s3Ae2}
        \begin{aligned}
        -E&=p_t-\frac{1}{2}g_{tt,r} S^{tr}
        \\
        J&=p_\varphi-\frac{1}{2}g_{\varphi\varphi,r} S^{\varphi r},
        \end{aligned}
\end{equation}
Moreover, since the particle's motion is constrained to the equatorial plane, the antisymmetric spin tensor $S^{\alpha \beta}$ has only two independent components, namely \cite{Abdulxamidov:2023jfq}:
\begin{equation}
        \label{s3Ae3}
        \begin{aligned}
        S^{tr}&=-\frac{p_\varphi s}{\sqrt{-g_{tt} g_{rr} g_{\varphi\varphi}}},\\
        S^{\varphi r}&=\frac{p_t s}{\sqrt{-g_{tt} g_{rr} g_{\varphi\varphi}}},
        \end{aligned}     
\end{equation}
where $s$ represents the specific angular momentum of the particle (spin).

Putting Eq.~(\ref{s3Ae3}) into Eq.~(\ref{s3Ae2}), the expressions for the energy $E$ and the total angular momentum $J$ can be rewritten as
\begin{equation}
\label{s3E3}
        \begin{aligned}
        -E&=p_t+\frac{s p_\varphi g_{tt,r} }{2\sqrt{-g_{tt} g_{rr} g_{\varphi\varphi}}}=p_t- \frac{s h'(r)}{2r\sqrt{h(r)/f(r)}} p_\varphi
        \\
        J&=p_\varphi-\frac{s p_t g_{\varphi\varphi,r}}{{2\sqrt{-g_{tt} g_{rr} g_{\varphi\varphi}}}}=p_\varphi-\frac{s  }{{\sqrt{h(r)/f(r)}}}p_t ,
        \end{aligned}
\end{equation}
 and solving the system for $p_t$ and $p_\phi$, we can easily obtain:
\begin{eqnarray}
     \label{s3Ae5}
p_t&=&\frac{-2rh(r)E+h'(r)\sqrt{h(r)/f(r)}Js}{2rh-h'(r)f(r)s^2} \, ,
\\ 
\label{s3Ae6}
p_\varphi&=&\frac{2rh(r)J-2r\sqrt{h(r)f(r)}sE}{2rh-h'(r)f(r)s^2} \, .
    \end{eqnarray}
   
Now, from Eq.~(\ref{s3e7}), we can get the expression for the radial canonical momentum as: 
\begin{equation}
\label{s3Ae7}
        (p^r)^2=-g^{rr}[g^{tt}p_t^2+g^{\varphi\varphi}p_\varphi^2+m^2].
\end{equation}
From $g^{\alpha\beta}=1/g_{\alpha\beta}$, we get $g^{tt}=-1/h(r)$, $g^{rr}=f(r)$ and $g^{\varphi\varphi}=1/r^2$ in the equatorial plane where $\theta=\pi/2$. By inserting the Eqs.(\ref{s3Ae5}-\ref{s3Ae6}) and the components of the metric tensor into the Eq.(\ref{s3Ae7}), it is not difficult to get the quadratic equation in terms of the energy $E$ of the spinning particle as given below:
\begin{equation}
    \label{s3Ae8}
  ( u^r)^2=\frac{\alpha \mathcal{E}^2+\delta \mathcal{E}+\gamma}{\rho},
    \end{equation}
where we have used simple relationship for the momenta $(u^r)^2=(p^r/m)^2$ and introduced  dimensionless quantities ($\mathcal{E}=E/m$, $\mathcal{L}=L/m$).

Here, new notations denote:
\begin{eqnarray}
    \label{notations}
\alpha&=&4h(r)f(r)[r^2-f(r)s^2],\\
\delta&=&4\mathcal{J}\sqrt{h(r)f(r)}f(r)s[2h(r)-rh'(r)],\\
\gamma&=&\mathcal{J}^2f(r)[f(r)h'(r)^2s^2-4h(r)^2]-f(r)\rho,\\
\rho&=&[2rh(r)-h'(r)f(r)s^2]^2.
\end{eqnarray}

We can now  rewrite Eq.~(\ref{s3Ae8}) in the form:
\begin{equation}
      (u^r)^2=\frac{\alpha}{\rho}(\mathcal{E}-V_+)(\mathcal{E}-V_-)\ .
\end{equation}
In order to get the circular motion of the spinning particles, we set the condition $p^r=0$, of which we can define the effective potential in the following form \cite{Abdulxamidov:2023jfq,2019PhRvD.100j4052T}:
\begin{equation}
      V_{\pm}=\frac{-\delta\pm\sqrt{\delta^2-4\alpha\gamma}}{2\alpha}\ .
\end{equation}
However, to satisfy the condition $(u_r)^2\geq0$, the specific energy of spinning particles must meet one of two following conditions: 
(i) $\mathcal{E}>V_+$ and 
(ii) $\mathcal{E}<V_-$ .

Assuming that spinning particles have positive energy (the first condition), we completely define the effective potential as $V_{eff}=V_+$:
\begin{equation}
    \label{s3Ae9}
     V_{eff}=\frac{-\delta+\sqrt{\delta^2-4\alpha\gamma}}{2\alpha} \, .
\end{equation}

Fig.~~\ref{veffs} demonstrates the radial dependence of the effective potential for different values of the parameters $q_1$, $q_2$, $s$ and $\sigma$ at fixed angular momentum $\mathcal{L}=3.0$ .
So, the left panel of the upper line of the Fig.~\ref{veffs} illustrates the behavior of the effective potential for various values of $q_1$ at fixed $q_2=0.5$. Whereas, the right panel of the upper line shows the behaviour of the effective potential for various values of $q_2$ at fixed  $q_1=1.0$. In both cases, $\sigma=0.8$ and the spin of the particle is positive and fixed $s=0.5$. Here, it can be observed that the increase in both parameters, $q_1$ and $q_2$, increases the value of the effective potential. 

On the other hand, the first graph of the lower line displays the radial dependence of the effective potential on various values of the spin of the particle at fixed $q_1=1.0$, $q_2=0.5$ and $\sigma=0.8$. Here we can notice that the value of the effective potential goes up when the spin of the particle is changing from the negative to the positive value. Last, the right panel of the lower line illustrates the radial dependence of the effective potential for different values of the $\sigma$, holding constant $q_1=q_2=0.5$ and $s=0.5$.  In contrast to all the graphs, the opposite behaviour of the effective potential can be seen by rising the values of the $\sigma$, namely when the value of the $\sigma$ increases the value of the effective potential decreases, but the shift in the value of the $V_{eff}$ is very small.

\subsection{Innermost stable circular 
orbit \label{sec4b}}

In this subsection we consider the innermost stable circular orbits (ISCO) of the spinning particle around Einstein-geometric Proca AdS black hole. Here we utilize the following three conditions in order to find the ISCO of the particle:
\begin{enumerate}
    \item $dr/d\tau=0$ or $\mathcal{E}=V_{eff}(r)$ \, ,
    \item $d^2r/d\tau^2=0$ or  $V_{eff}'(r)=0$ \, ,
    \item $d^2V_{eff}/dr^2\geq0$.
\end{enumerate}
Obviously, the first condition is necessary for the motion with constant radius, i.e. there is no change in the radial motion over time. Consequently, it leads to the equality of the energy and the effective potential. 
In addition, the second condition indicates the absence of acceleration, the motion with a constant speed. These two conditions should be met to get the circular orbits for a particle. However, the inner orbits and their stability are achieved by satisfying the last condition. Thus, the satisfaction of all three conditions simultaneously leads to the movement of the particle in the trajectory of the innermost stable circular orbit. 

\begin{figure*}[t]
    \begin{center}
    \includegraphics[scale=0.45]{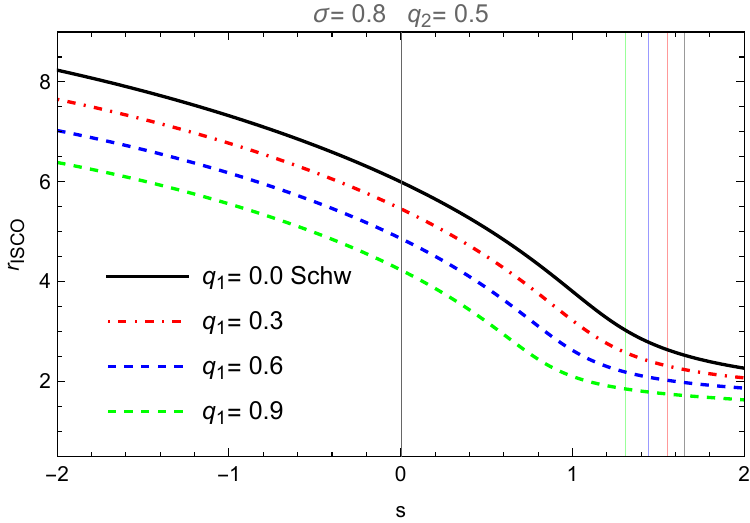}
    \includegraphics[scale=0.45]{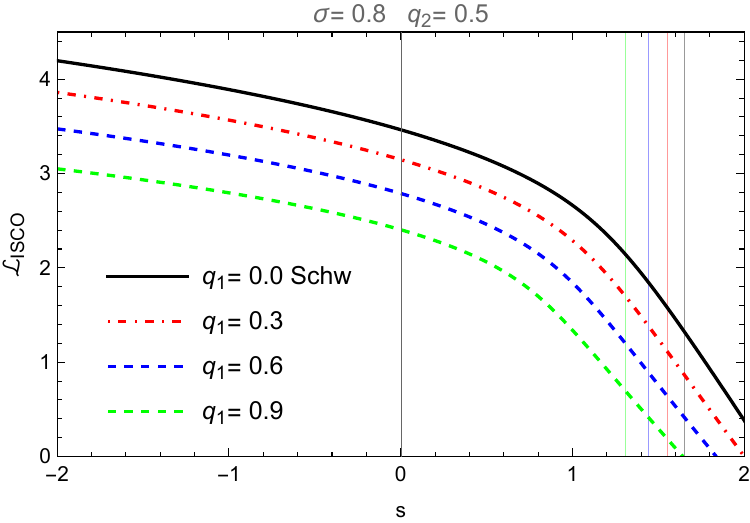}
    \includegraphics[scale=0.45]{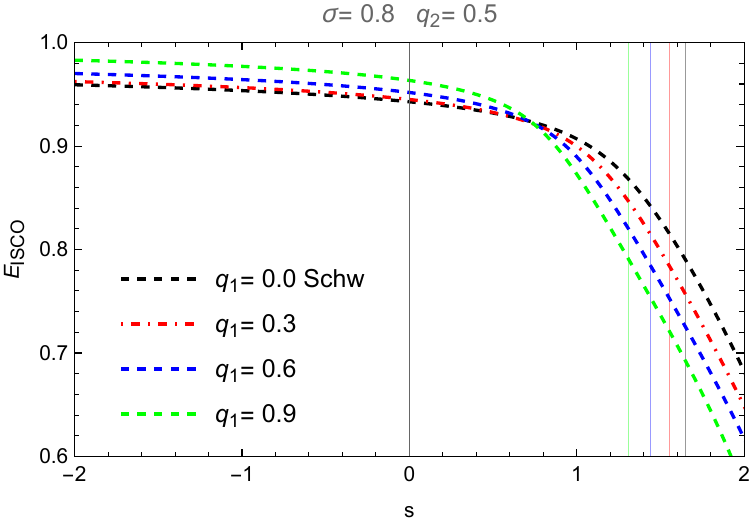}
    \caption{The variation of the radius, specific angular momentum, and energy at ISCO in spin $s$ for different values of $q_1$ at fixed $q_2=0.5$ and  $\sigma=0.8$, respectively. Here,  straight colorful lines in the graph define the superluminal boundary of the particle's spin for each value of the $q_1$, respectively. The left side of the lines corresponds to the time-like particles, while the right side - space-like particles. }
     \label{ISCOspin}
    \end{center}
\end{figure*}

\begin{figure*}[t]
    \begin{center}
    \includegraphics[scale=0.45]{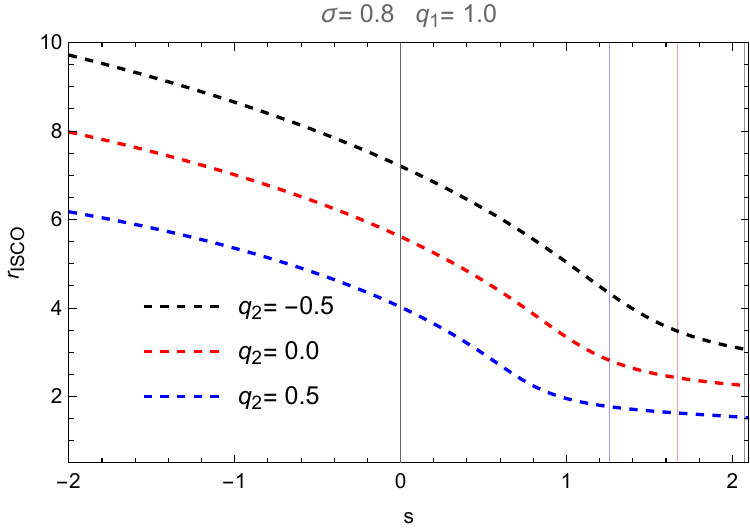}
    \includegraphics[scale=0.45]{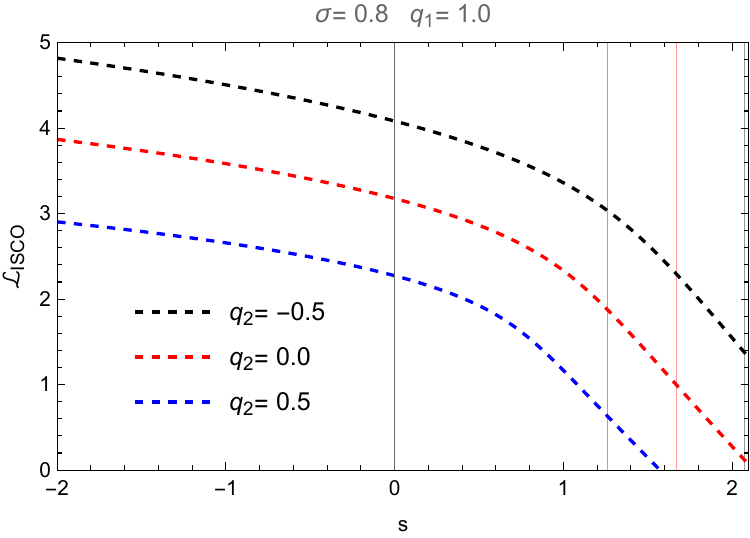}
     \includegraphics[scale=0.45]{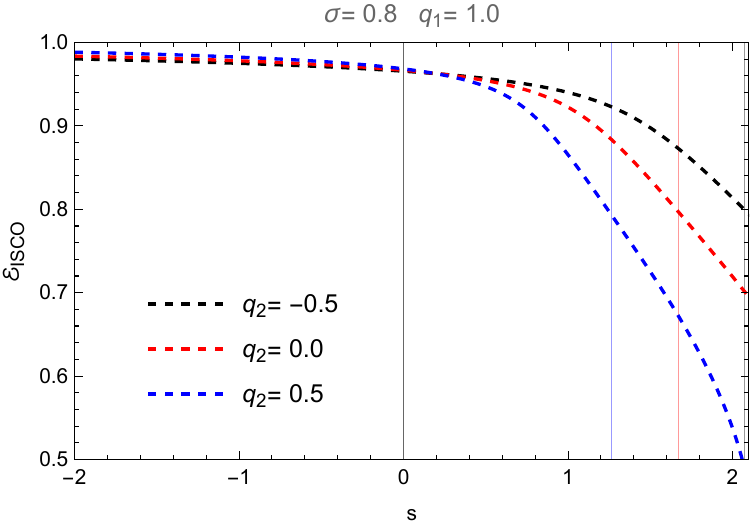}
    \caption{The variation of the radius, specific angular momentum, and energy at ISCO with respect to spin $s$ for various values of $q_2$ and for $q_1=1.0$ and  $\sigma=0.8$. The vertical colorful lines define the superluminal constraint on the spin for each value of $q_2$ in the graph.}
\label{ISCOspin2}
    \end{center}
\end{figure*}

\begin{figure*}[t]
    \begin{center}
    \includegraphics[scale=0.5]{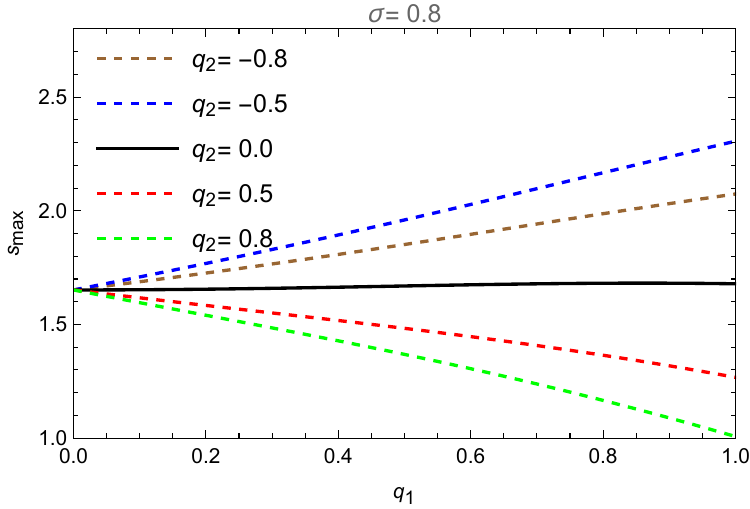}
    \includegraphics[scale=0.5]{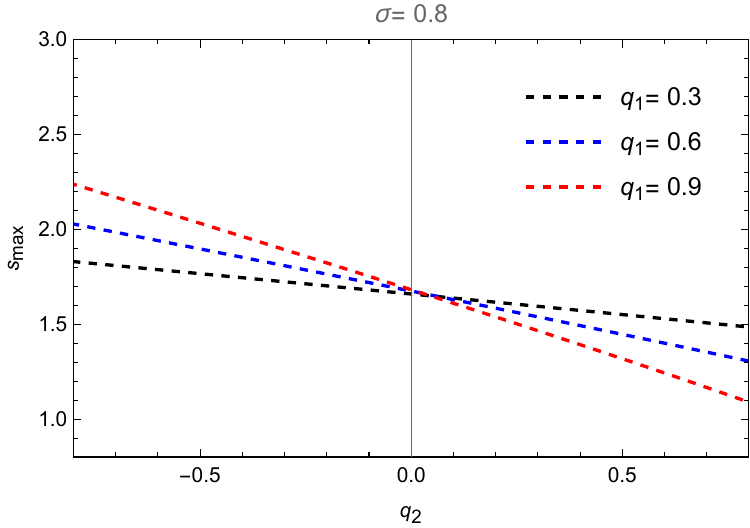}
    \caption{The dependence of the critical values of spin $s_{max}$ on parameter $q_1$ for various values of $q_2$ (left panel) and on parameter $q_2$ for various values of $q_1$ (right panel) for $\sigma=0.8$.}
    \label{scr}
    \end{center}
\end{figure*}

\begin{figure*}[t]
    \begin{center}
    \includegraphics[scale=0.5]{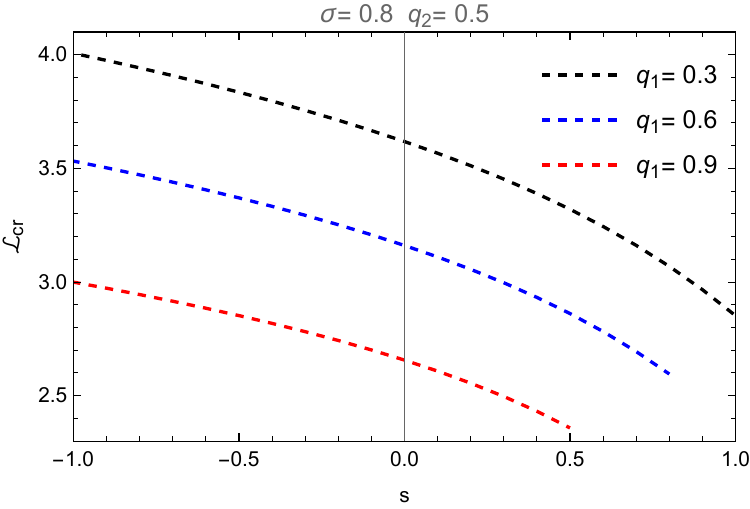}
    \includegraphics[scale=0.5]{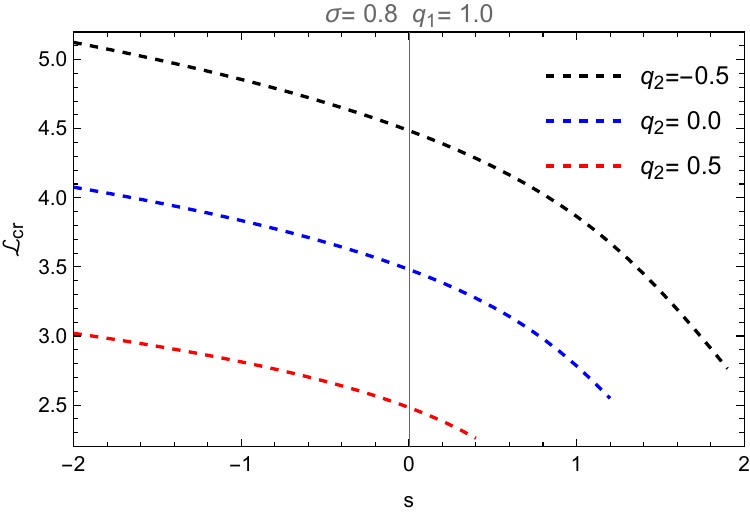}
    
    \caption{The dependence of the critical angular momentum on the particle's spin $s$ for various values of parameters $q_1$ and $q_2$ for $\sigma=0.8$.}  
    \label{Lcrs} 
    \end{center}
\end{figure*}

\begin{figure*}[t]
    \begin{center}
\includegraphics[scale=0.45]{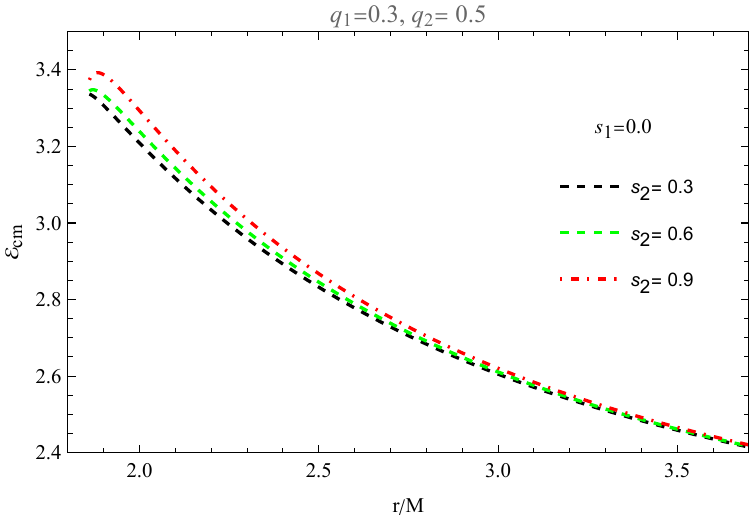}
\includegraphics[scale=0.45]{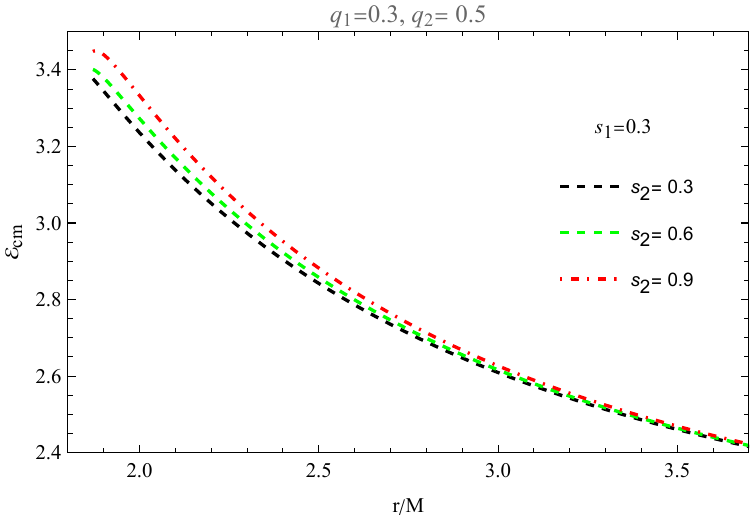}
\includegraphics[scale=0.45]{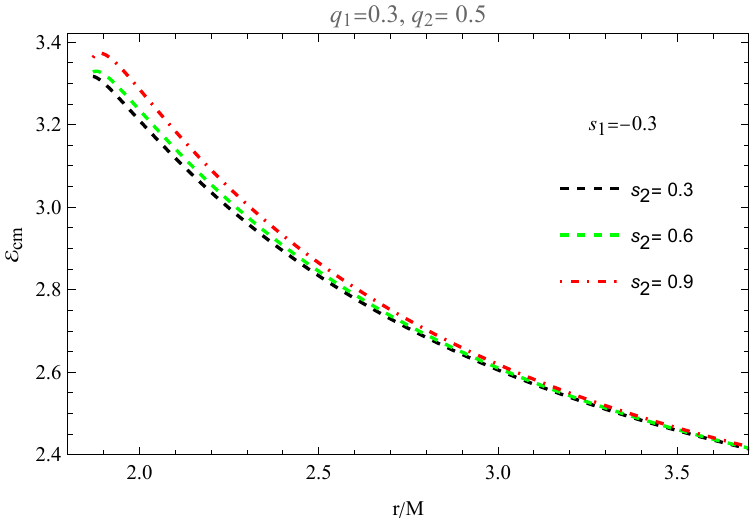}
\includegraphics[scale=0.45]{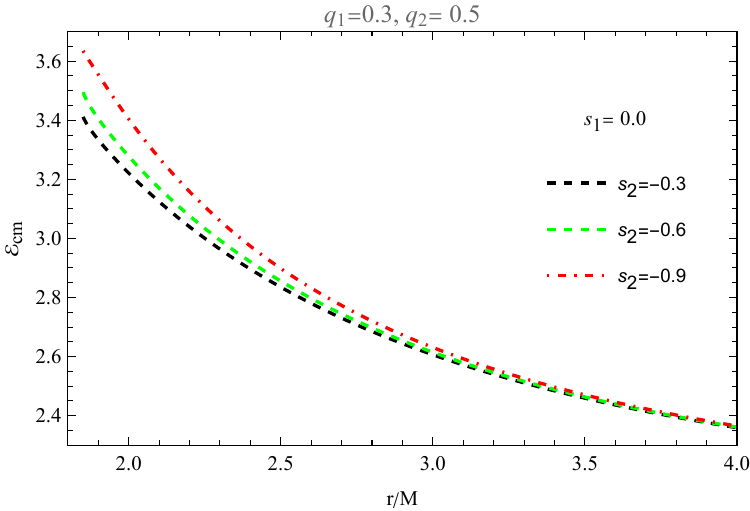}
\includegraphics[scale=0.45]{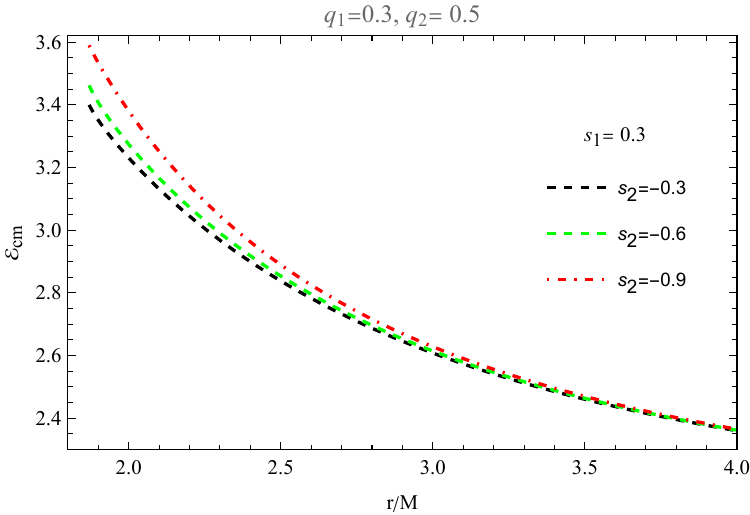}
\includegraphics[scale=0.45]{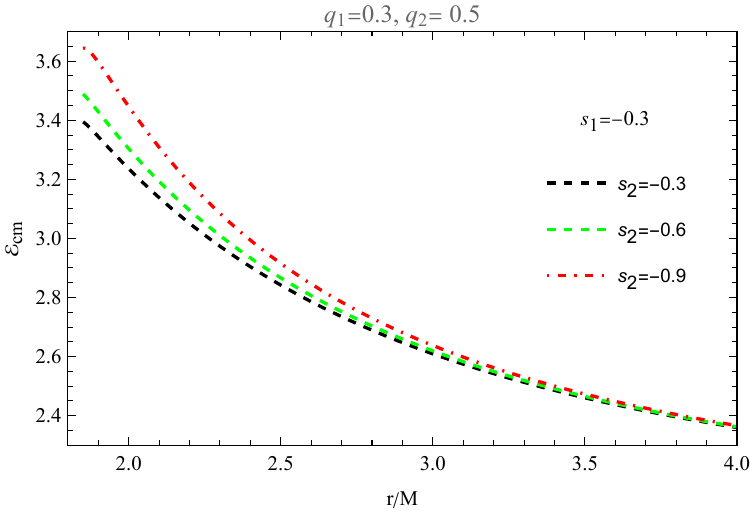}
 \caption{The radial dependence of the center-of-mass energy of spinning particles for various values of the spin for $q_1=0.3$, $q_2=0.5$ and $\sigma=0.8$. The figure contains six panels. All panels of the first line illustrate the center-of-mass energy for the various positive values of the second particles spin for different fixed values of the spin of the first one. The second line displays the opposite - the center-of-mass energy for the fixed values of the first and various negative values of the spin of the second particle.}
 \label{Ecmspin}
    \end{center}
\end{figure*}

Figs.~\ref{ISCOspin},~\ref{ISCOspin2} illustrate the dependence of radius, specific angular momentum and specific energy at ISCO ($r_{ISCO},\mathcal{L}_{ISCO}, \mathcal{E}_{ISCO}$) on spin $s$ of the particle for various values of parameters $q_1$ and $q_2$, respectively. We analyse below each plot separately to see the behaviour of the functions and differences with respect to the model parameters.

Fig.~\ref{ISCOspin} displays the spin-dependence of the radius, specific angular momentum and  specific energy at ISCO for various values of $q_1$ for $q_2=0.5$ and $\sigma=0.8$.   
Note that we have included here the Schwarzschild condition $q_1=0$ to compare and be convinced of the correctness of the results. It is observed that the rise in the parameter $q_1$ decreases both $r_{ISCO}$ and $\mathcal{L}_{ISCO}$ in the first two graphs of the Fig.~\ref{ISCOspin}. Furthermore, one can notice that the specific energy at ISCO (the last graph of the plot) slightly increases when $q_1$ increases for particles with negative spin and positive spin  until $s\approx+0.72$ (the specific value of the particle's spin, joining point of the lines). Afterwards, the behaviour of $\mathcal{E}_{ISCO}$ is similar with $r_{ISCO}$ and $\mathcal{L}_{ISCO}$ i.e., when the value of the $q_1$ increases, the value of the specific energy decreases. Noticeably, all functions ($r_{ISCO},\mathcal{L}_{ISCO},\mathcal{E}_{ISCO}$) diminish with the rise of the value of the particle's spin. 
Additionally, one can see that all graphs have straight colorful lines corresponding to lines of the functions for each value of the $q_1$, which define the superluminal boundary for the particle's spin. As have already mentioned in Sec.~\ref{sec3b}, the trajectory of the particle can be space-like for some values of the spin. So, the left side of the graph corresponds to the time-like particles, while the right side - space-like ones, and the colorful vertical lines denote the boundary for the spin of the particle.

Fig.~\ref{ISCOspin2} represents the dependence of the radius, specific angular momentum, and specific energy of the ISCO on the spin $s$ for the different values of $q_2$ at fixed  $q_1=1.0$ and $\sigma=0.8$. One can notice that the graphs have a similar trend as in Fig.~\ref{ISCOspin}, when $q_2$ increases the radius, specific angular momentum and specific energy at ISCO decreases. Yet, the shift on the values of the functions  is much more significant than the one in the Fig.~\ref{ISCOspin}. In addition, there is also a joining point of the lines in the plot of the specific energy at ISCO, at around $s=+0.2$, after which the lines start to differ noticeably. The Fig.~\ref{ISCOspin2} has also colorful vertical lines for each value of the $q_2$, representing the superluminal constraint for the particle's spin. 

The variation of the critical values of the spin of $s_{max}$ in parameters $q_1$ and $q_2$ is presented in the Fig.~\ref{scr}. The left panel of the Fig.~\ref{scr} shows the dependence of the critical values of spin on $q_1$ for different values of the parameter $q_2$ when the parameter $\sigma=0.8$ is constant for all cases. It should be noted that there are two totally different trends for negative and positive values of $q_2$: a rise in the negative values of $q_2$ results in the increase of the critical values of the particle's spin and is getting bigger with the rise of  $q_1$, while for positive values of the $q_2$ correspond small values of $s_{max}$ and it decreases with the rise of $q_1$. A special case is represented when $q_2=0$, namely the critical value of the particle's spin is almost constant even if the values of parameter $q_1$ go up (here the critical values of the spin change from $s_{max}=1.65181$ to $s_{max}=1.68186$ when $q_1=0\div1$).

The right panel of the Fig.~\ref{scr} illustrates the behaviour of the critical values of the spin as a function of the parameter $q_2$ for different values of $q_1$. The symmetric behaviour of the $s_{max}$ in terms of negative and positive values of $q_2$ can be observed from the graph. So, when $q_2$ is negative the values of $s_{max}$ go up with the rise of $q_1$ and, vice versa, the values of $s_{max}$ go down when $q_1$ increases at positive values of $q_2$. It should be mentioned that with the rise of $q_2$ from negative to positive values the critical values of the particle's spin decrease all the time. Here, again, we can notice the point where the value of $s_{max}$ is almost the same for different values of $q_1$.

\section{Collision of spinning particles\label{sec5}}

This section is devoted to investigating the head-on collision of two particles near the horizon of the black hole and the ultra-high energy produced by it. Here, we assume that particles coming from infinity fall freely and produce a head-on collision near the horizon of the black hole. For this aim, we first analyze the angular momentum of the particle, which is extremely important. 

\subsection{Angular momentum}
It is worth mentioning that to produce a head-on collision of two particles, we need to know the critical values of the angular momentum of the particle for which the collision is possible. In order to find these values of the angular momentum $L_{cr}$, the following conditions should be satisfied:

\begin{itemize}
    \item [i)] $\dot{r}^2=0$\, ,
    \item[ii)] $d\dot{r}^2/dr=0$ \, .
\end{itemize}
Obviously, the particles with higher angular momentum than the critical values cannot approach the black hole and produce a collision.

Fig.~\ref{Lcrs} demonstrates the critical angular momentum of a spinning particle as a function of spin $s$ for different values of parameters $q_1$ and $q_2$ at fixed $\sigma=0.8$. The left panel of the plot shows the behavoir of the $L_{cr}$ for different values of $q_1$ when $q_2=0.5$ is constant. One can see from the graph, the values of the critical angular momentum decrease noticeably when $q_1$ increases. Note, that the values of the  critical angular momentum moderately go down with the increase of the values of the particle's spin and the higher the value of the $q_1$ the lower the value of the $L_{cr}$ and spin $s$ of the particle.

In addition, the right panel of the Fig.~\ref{Lcrs}
illustrates the dependence of the critical angular momentum on spin of the particle for different values of $q_2$ at fixed $q_1=1.0$. As can be seen, the values of $L_{cr}$ drop considerably with the increase of $q_2$, as in the case of $q_1$. It should be noted that both the spin and the critical angular momentum of the particle have higher values when $q_2$ is negative.

\subsection{The center-of-mass energy of spinning particles}
Now, we utilize the following expression for computing the center-of-mass energy $E_{cm}$ of the colliding spinning particles \cite{Abdulxamidov:2023jfq}:
\begin{eqnarray}
        E_{cm}^2&=&-g^{\mu\nu}(p^{(1)}_{\mu}+p^{(2)}_{\mu})(p^{(1)}_{\nu}+p^{(2)}_{\nu}) \nonumber \\ &=& m_1^2+m_2^2-2g^{\mu\nu}p^{(1)}_{\mu}p^{(2)}_{\nu},
\end{eqnarray}
where $p^{(1)}_{\mu}$ and $p^{(2)}_{\mu}$ are four-momentum of spinning particles, respectively (see Eqs.~\ref{s3Ae5} and \ref{s3Ae7}). 
Here, we assume that particles have the same mass $m_1=m_2=m$, but different four-momenta $p^{(1)}_{\mu}$ and $p^{(2)}_{\mu}$. Thus, the final expression for the center-of-mass energy of colliding particles has the form:
\begin{eqnarray}
    \mathcal{E}_{cm}^2&=&\frac{E_{cm}^2}{2m^2}\nonumber \\&=&1-\frac{g^{tt} p^{(1)}_t p^{(2)}_t+g^{rr} p^{(1)}_r p^{(2)}_r+g^{\varphi\varphi} p^{(1)}_\varphi p^{(2)}_\varphi}{m^2}\ .\ \ 
\end{eqnarray}
\\
Fig.~\ref{Ecmspin} displays the radial dependence of the center-of-mass energy $E_{cm}$ of colliding spinning particles for different values of spin $s$ at fixed $q_1=0.3$ and $q_2=0.5$. Here, all calculations have been done at fixed $\sigma=0.8$ and $L=2.5$. It
consists of six panels, divided into two rows.

One can see that the first line of the plot consists of three graphs, each of which has a constant  value of the spin of the first particle with varying positive values of the spin of the second particle. Noticeably, that by increasing the value of the spin of the second particle, the value of the center-of-mass energy $E_{cm}$ also increases, but only slightly. 
On the other hand, the graphs of the  second line show the behaviour of the center-of-mass energy $E_{cm}$ for constant value of the spin of the first with different negative values of the spin of the second particle. Similarly with the graphs of the first line, the rise in the negative values of the second particle's spin the $E_{cm}$ energy gets higher value, where the shift in the value is noticeable.

Considering all panels, we can see that the maximum values of $E_{cm}$ are released during a collision when the first particle is spinless ($s_1=0$) or with a negative spin ($s_1=-0.3$) and the second particle has a negative spin ($s_2=-0.3;-0.6;-0.9$), and the higher the negative spin of the second particle is, the more energy is released during the collision. In the meantime, the minimum center-of-mass energy $E_{cm}$ corresponds to the case when the spin of the first particle is negative ($s_1=-0.3$) and the second particle has positive values of the spin ($s_2=0.3;0.6;0.9$). Note, the smaller the value of the spin of the second particle, the less energy $E_{cm}$ is released during the collision.

\section{Conclusions
\label{Sec:conclusion}}

In this work, we have conducted a detailed investigation into the dynamics of spinning test particles orbiting Einstein-geometric Proca Anti-de Sitter (AdS) compact objects. By solving the Mathisson-Papapetrou-Dixon (MPD) equations under the Tulczyjew spin-supplementary condition, we have systematically analyzed how the interplay between particle spin and the novel parameters of the extended metric-Palatini gravity model influences orbital behavior, stability, and high-energy collision processes.

Our principal findings can be summarized as follows:

\begin{itemize}
    \item  The parameters \( q_1 \) and \( q_2 \), characterizing the geometric Proca field, exert a significant influence on the innermost stable circular orbits (ISCO). An increase in either parameter leads to a reduction in the ISCO radius, angular momentum (\( \mathcal{L}_{ISCO} \)), and energy (\( \mathcal{E}_{ISCO} \)). This indicates a stronger effective gravitational attraction compared to a Schwarzschild-AdS black hole. The particle's spin \( s \) further modulates this structure, with positive spin generally increasing orbital energy and negative spin enhancing the inward shift of the ISCO.

    \item We identified critical spin values \( s_{max} \) that demarcate the boundary between physical (time-like) and non-physical (space-like) trajectories. The parameter space for physically allowed spins is highly sensitive to the Proca charges. Notably, negative values of \( q_2 \) permit a wider range of spin values, including higher positive spins, whereas positive \( q_2 \) values significantly restrict the maximum allowable spin.

    \item The analysis of head-on collisions near the horizon reveals that Einstein-geometric Proca AdS compact objects can act as potent particle accelerators. The center-of-mass energy \( \mathcal{E}_{cm} \) is substantially enhanced when the colliding particles have aligned negative spins. The critical angular momentum \( \mathcal{L}_{cr} \), necessary for particles to reach the horizon, decreases with increasing \( q_1 \) and \( q_2 \), facilitating more extreme collisions for a broader range of orbital parameters.
\end{itemize}

In summary, our results demonstrate that the dynamics of spinning particles serve as a sensitive probe of the spacetime geometry governed by the Einstein-geometric Proca theory. The distinctive modifications induced by the Proca parameters \( (q_1, q_2, \sigma) \) manifest in the ISCO properties, superluminal bounds, and collision energies provide clear, observable signatures that distinguish these compact objects from their General Relativity counterparts. These findings pave the way for future explorations, including the dynamics around rotating versions of these black holes and the implications of these results for astrophysical observations in the strong-gravity regime.

\section*{Acknowledgements}
B. P. would like to acknowledge networking support by the COST Action CA21106 - COSMIC WISPers in the Dark Universe: Theory, astrophysics, and experiments (CosmicWISPers), the COST Action CA23115 - Relativistic Quantum Information (RQI), and the COST Action CA23130 - Bridging high and low energies in search of quantum gravity (BridgeQG) funded by COST (European Cooperation in Science and Technology).

\bibliographystyle{spphys}


\end{document}